\documentclass{nime-alternate} 

\usepackage{anonymize} 

\begin{document}
\conferenceinfo{NIME'19,}{June 3-6, 2019, Federal University of Rio Grande do Sul, ~~~~~~  Porto Alegre,  Brazil.}
\title{Learning from History: \\ Recreating and Repurposing Sister Harriet Padberg's \\ Computer Composed Canon and Free Fugue}
\numberofauthors{6} 
\author{
%
%
\alignauthor
\anonymize{Richard Savery}\\
       \affaddr{\anonymize{Georgia Tech Center for Music Technology}}\\
       \email{\anonymize{rsavery3@gatech.edu}}
\alignauthor
\anonymize{Benjamin Genchel}\\
       \affaddr{\anonymize{Georgia Tech Center for Music Technology}}\\
       \email{\anonymize{benjiegenchel@gmail.com}}
\alignauthor \anonymize{Jason Smith}\\
       \affaddr{\anonymize{Georgia Tech Center for Music Technology}}\\
       \email{\anonymize{jsmith775@gatech.edu}}
\and  
\alignauthor \anonymize{Anthony Caulkins}\\
       \email{\anonymize{anthonycaulkinsmusic@gmail.com}}
\alignauthor \anonymize{Molly Jones}\\  
       \email{\anonymize{mollyej@uci.edu}}
\alignauthor \anonymize{Anna Savery}\\
       \email{\anonymize{anna@annasavery.com}}
}

\date{19 January 2019}

\maketitle
\begin{abstract}

Harriet Padberg wrote \textit{Computer-Composed Canon and Free Fugue} as part of her 1964 dissertation in Mathematics and Music at Saint Louis University. This program is one of the earliest examples of text-to-music software and algorithmic composition, which are areas of great interest in the present-day field of music technology. This paper aims to analyze the technological innovation, aesthetic design process, and impact of Harriet Padberg's original 1964 thesis as well as the design of a modern recreation and utilization, in order to gain insight to the nature of revisiting older works.
Here, we present our open source recreation of Padberg's program with a modern interface and, through its use as an artistic tool by three composers, show how historical works can be effectively used for new creative purposes in contemporary contexts. \textit{Not Even One} by \anonymize{Molly Jones} draws on the historical and social significance of Harriet Padberg through using her program in a piece about the lack of representation of women judges in composition competitions. \textit{Brevity} by \anonymize{Anna Savery} utilizes the original software design as a composition tool, and \textit{The Padberg Piano} by \anonymize{Anthony Caulkins} uses the melodic generation of the original to create a software instrument.

\end{abstract}





\section{Introduction}

In 1964, Harriet Padberg completed the first dissertation on algorithmic composition using a computer, \textit{Computer Composed Canon and Free Fugue}, which converted text inputs into melodies using a custom algorithm. This work stood out at the time for using a text-to-music mapping instead of the random number generation based systems used in her contemporaries' pieces, e.g. \textit{Illiac Suite} by Lejaren Hiller\cite{fernandez2013}.

It received moderate attention in the years following its creation, including a 1968 sonic realization by Max Matthews \cite{ariza2011} and mention in a 1970 survey of computer-generated music by Lejaren Hiller \cite{hiller1968}. However, in the present, her work has largely been forgotten; it is not commonly known, nor spoken about outside of select papers, and a single recreation of the fugue according to the score tables and code included in the dissertation was published accompanying Computer Music Journal, vol. 35, no. 4. The delay of its inclusion until a 2011 issue, 47 years after the dissertation was published \cite{nilson2011}, presents not only a lack of attention towards a significant historical work but also the loss of opportunity to learn from and adapt it for new musical creations. 

In this paper, we present a history of \textit{Computer Composed Canon and Free Fugue}, describe its underlying technology and aesthetic characteristics, and discusses its broader impact, or lack thereof. We also present an  open-sourced recreation of the algorithm, implemented in Python 3.7\footnote{https://www.python.org/, last accessed: 19th Apr 2019}, alongside three examples of its use in compositions by modern composers. Each use represents a general mode by which older technology can be repurposed into a new context with new meaning.

The first composition, \textit{Not Even One}, by \anonymize{Molly Jones}, shows how the context around a historical piece of technology can act as a lens through which we can view contemporary issues, as well as confer contextual meaning onto its outputs. \textit{Brevity}, by \anonymize{Anna Savery}, shows how the function of a piece of technology, independent of context, can be useful for imbibing work with hidden meaning, and how the particular aesthetic of an older work can still be desireable long past its deprecation. Finally, \anonymize{The Padberg Piano} by \anonymize{Anthony Caulkins} shows how key concepts from older works still serve as wells of inspiration, and can inspire creative technical adaptations such as the conception of a new tuning system. 

We contend that recreations of historical works allow for better understanding of technical and aesthetic practices of the past. Using these works in a modern context is fruitful both for creators and listeners concepts, sounds and ideas are revisited. The historical re-contextualization of work also allows new understandings of current work and climate through comparison to past creations.  

\section{Original Work}

\subsection{Background}
Harriet Padberg (November 13, 1922 - January 2, 2014) spent her life as a member of the ministry of a Roman Catholic women's congregation titled the Society of the Sacred Heart. She received a PhD in mathematics and music from Saint Louis University, after which she became a professor at Maryville University in St. Louis. Padberg's PhD dissertation, \textit{Computer-Composed Canon and Free-Fugue}, written in 1964, is likely the first dissertation on computer algorithmic composition. In this work, Padberg combined traditional music theory methods with a novel text-to-music algorithm, contrasting the stochastic approaches that were widely used at the time.

During her following tenure at Maryville, she would further advance the use of computer music through her establishment of the university's Music Therapy program alongside Ruth Sheehan \cite{tenYearsMaryville}, paving the way for the creation of its Occupational Therapy and Speech-Language Pathology programs \cite{maryvilleUniversity} a few years later. The Music Therapy program would steadily expand during its first ten years, including the creation of a work study program allowing students to gain professional experience in the field while studying\cite{tenYearsMaryville}. Padberg herself worked as a therapist as well, and additionally went on to begin her own music ministry in 1992 \cite{rscjWebsite}.

\subsection{Technology}
\textit{Computer Composed Canon} centers around an algorithm written in FORTRAN \cite{padberg1964} which mapped text to pitch, rhythm and musical phrasing. 

The first part of the algorithm reads in up to 160 characters of text \cite{padberg1964, bashe1986}, and maps each constituent letter to an integer which indexes a length 24 table of frequencies consisting of partials 24 through 47 of the overtone series for some root note. The letter Y is mapped to the letter I on its first appearance, and to the letter Z for subsequent appearances. Additionally, the letters W and V are assigned the same frequency. During this same pass, the input is divided into blocks containing words of five or more letters each, and words with less than five letters adjoined with neighboring words to form longer blocks. Each letter is assigned a block number, and an index into their assigned block \cite{padberg1964}. 

Once this mapping is determined, a second pass is performed during which a number of other meta-data is extracted from the text, including the number of vowels and consonants per block, the total number of vowels and consonants for the entire phrase, and the number of notes in each octave. The set of blocks, starting from the beginning of the input phrase, whose constituent letters sum to less than or equal to 24 are used as a ``tone row'', as well as the melody line of the Canon. Repeating letters in the tone row are mapped to the same note in a different octaves according to the number of repetitions and whether that number of repetitions is odd or even, making each pitch in the ``tone row'' unique. The rhythm of the melody is derived via a process that takes into account the relative primality of the length of the tone row, and the number of vowels, consonants and blocks it contains. 

The algorithm outputs the composition on a punch card with two columns, one of which specifies the melody in a piano roll format with time divided into ``ticks'' representing a constant time span. Note lengths are specified by number of ticks in which the note is a ``on''. The other denotes ``the number of units of the rhythmic pattern''. The melody line is repeatedly written to the cards in an overlapping fashion, with each successive line beginning one measure after the beginning of the preceding line \cite{padberg1964}. 

The original program was written for and run on the IBM 1620, which was an inexpensive scientific computer that could store 20-60 thousand digits at a time. Due to space constraints, the algorithm was eventually moved to an IBM 7072, where it underwent some technical changes to generalize its process such that it could generate not only Canons, but ``Free-Fugues'' for two or three voices as well \cite{bashe1986}. 

Though Padberg's algorithm is surely the first composition technologically realized specifically on the IBM 1620 and 7072, it is predated by other uses of computers for algorithmic composition. David Caplin and Dietrich Prinz ran some of the very first experiments composing music with computers in 1955. They created an implementation of Mozart's dice game among other probabilistic music programs \cite{ariza2011}. A few years later, Martin Klein and Douglas Bolitho created a program called \textit{Push Button Bertha} which randomly output numbers, representative of chromatic pitch classes, and structured them using a set of hard-coded rules \cite{hiller1968}. Hiller and Isaacson's \textit{Illiac Suite} preceded Padberg's work by seven years. However, their work sought to use the computer to produce notes and phrases which could be arranged by musicians, whereas Padberg's algorithm is meant to produce complete pieces of music on its own \cite{ariza2011}. Milton Babbit's work with the RCA Mark II similarly used the computer in composition for its precision and mathematical translation abilities. Babbit's serialism features greatly throughout Padberg's thesis, providing clear influence for many of her algorithmic and aesthetic choices \cite{padberg1964}.

\subsection{Aesthetics}
The aesthetics of \textit{Computer Composed Canon and Free Fugue}, reflecting the aesthetics of Padberg herself at the time, are rooted in Western Classical music tradition, particularly 20th-century Serialism are described within. Padberg's interest in Serialism was rooted in her educational and cultural backgrounds: she completed Masters degrees in both Musicology and Mathematics. As a devout Catholic, she was fond of the ancient Greek philosophies around the associations between mathematics, music, and the heavens. In her dissertation, Padberg describes Serialism as a method by which to ``re-discover'' from Pythagoras the inherent relation of music and mathematical models. Her choice of the name, ``Canon'' to describe the piece is derived from use of the term in music literature, wherein it typically refers to a melody which reoccurs at a set interval underneath itself; a common technique in 12-tone music. The term ``free-fugue'' however is not used literally. Instead, she uses the term to refer to multiple melodies appearing and crossing over one another\cite{padberg1964}. 

Her decision to map letters to pitches draws on J.S. Bach's \textit{Unfinished Fugue}, in which Bach famously built his own name into musical text, as well as later tributes to this work by Anton Webern, Robert Schumann, and Arnold Schoenberg\cite{padberg1964}. These works were created using a direct mapping in German musical notation - A to A, B to Bb, C to C, onwards, with H mapping to B Natural. These text to music mappings vary from those used by Padberg, who did not associate pitch name directly with its text representation with the exception of A natural, which was mapped to 440 Hz. Padberg chose to use a 24-microtonal scale, largely based on the relation between characters in the alphabet and the capability of controlling with software, yet does not state any specific composer or work as the inspiration. 

Padberg also drew inspiration from other past experiments in algorithmic composition. Influenced by the \textit{Illiac Suite}, she also drew on the work of John R. Pierce and M.E. Shannon from Bell Labs and Richard C. Pinkerton at the University of Florida. From Bell Labs she describes three chance pieces, however criticizes them as ``nothing very interesting either because the music is too surprising, too random, too unpredictable''\cite{padberg1964}.

\begin{table}[]
\centering
\begin{tabular}{ll}
\hline
\multicolumn{1}{|l|}{Letter} & \multicolumn{1}{l|}{Frequency (Hz)} \\ \hline
A                            & 440                                 \\
B                            & 458.33                              \\
C                            & 476.66                              \\
D                            & 495                                 \\
E                            & 513.33                              \\
F                            & 531.66                              \\
G                            & 550                                 \\
H                            & 568.33                              \\
I/Y                          & 586.66                              \\
J                            & 605                                 \\
K                            & 623.33                              \\
L                            & 641.66                              \\
\end{tabular}
\quad
\begin{tabular}{ll}
\hline
\multicolumn{1}{|l|}{Letter} & \multicolumn{1}{l|}{Frequency (Hz)} \\ \hline
M                            & 660                                 \\
N                            & 678.33                              \\
O                            & 696.66                              \\
P                            & 715                                 \\
Q                            & 733.33                              \\
R                            & 751.66                              \\
S                            & 770                                 \\
T                            & 788.33                              \\
U                            & 806.66                              \\
V/W                          & 825                                 \\
X                            & 843.33                              \\
Y/Z                          & 861.66                             
\end{tabular}
\caption{Padberg's Letter to Frequency Mapping}
\end{table}

\section{PyPadberg}
PyPadberg\footnote{\anonymize{https://github.com/bgenchel/pypadberg, last accessed: 19th Apr 2019}} is an open-source recreation and interpretation of Padberg's dissertation written in Python. The program contains modules for parsing text, mapping frequencies and calculating rhythms as well as synthesizing resultant melodies. PyPadberg contains a full screen ASCII interface which runs in terminal. The remainder of this section describes the program and its implementation in detail.

\subsection{Synthesis}
To allow users to hear back and export audio of created melodies we included a simple synthesis engine. The engine uses pitch shifting and time-stretching to act as a simple sample player. Three wave files are included with the program: a violin sound, a sine-wave based synth sound, and an 8-bit sound. Users are encouraged to add their own files into the sample engine. 

\subsection{Interface}
We packaged our implementation in a python application with a full screen ASCII user interface that runs in terminal. The interface was created using the Asciimatics library \footnote{https://github.com/peterbrittain/asciimatics, last accessed: 19th Apr 2019}, which provides convenient access to low level console functionality and high level APIs for creating animations and widgets, and implementing various types of interactivity. 

We specifically chose to develop an ASCII terminal based interface in order to emulate the simplicity and minimalism of older computer interfaces. Admittedly, this aesthetic is misplaced in this context, as Padberg's program predated digital interfaces of any kind. At the time, Padberg did not even have the ability to hear her own outputs; her program output results in the form of punch cards. We had considered creating a visual element out of faith to the original work, though in the end concluded that it would have been difficult and impractical. 

Our interface consists of four screens - a splash screen, a text input screen, a processing screen, and a rendering screen, each of which is described below.

\subsubsection{Splash Screen and Text Entry}
The splash screen, which appears when the application is launched,  displays the application title ``PyPadberg'' in large block letters over an image. The image is ``Woman Operating an IBM 1620 Data Processing System,'' from the \textit{DeCarlo Photograph Collection}, which is in reference to the IBM 1620 computer on which Padberg originally implemented her algorithm.

The text entry screen consists of a short introduction to the application followed by a large text box in which a user can enter text. Our application departs from the original described by Padberg in that we do not limit the entry to 60 characters, which was simply a limitation of the IBM 1620 rather than a creative choice by Padberg. Buttons at the bottom of the screen allow the user to either submit their entered text for processing or quit the application.

\subsubsection{Processing}
After the user submits their text on the text-entry screen, it is processed by our algorithm. Each major step of the text processing part of the algorithm generates a log which is saved in a list, and then displayed on the processing screen. The processing screen (Fig. 1) contains a scrollable list displaying the processing logs, giving the user a window into some of the inner-workings of the program. It displays to what frequency and rhythm value each character is mapped. A ``Continue'' button at the bottom of the screen allows the user to move on to the next screen once they have had a chance to look over the logs. 

\subsubsection{Rendering}
The rendering screen allows the user to select options for rendering the processed text into audio, including selecting a sample instrument for the melody, and a number of voices if the user would like to play the melody back as a psuedo-fugue or psuedo-canon. A playback option, which plays the audio according to the selected options in-terminal, is given at the bottom of the panel. The generic names of the samples (one, two and three) allow the user to seamlessly swap in their own custom samples if they so desire by renaming the sample as any of those numbers and placing it in the \textit{src/assets/audio} subdirectory of the application.

Following the playback options, a text box is provided for the user to enter a name for saving files, with options provided at the bottom of the panel. The user is given the option to save their melody as a CSV file for use with Max/MSP, or as a wav file. When saving a CSV, regardless of the number of voices selected, only the monophonic melody derived from the text is saved.

The final two buttons in the bottom panel allow the user to return to the text-entry screen or quit the application.

\begin{figure}[]
	\centering
		\includegraphics[width=1\columnwidth]{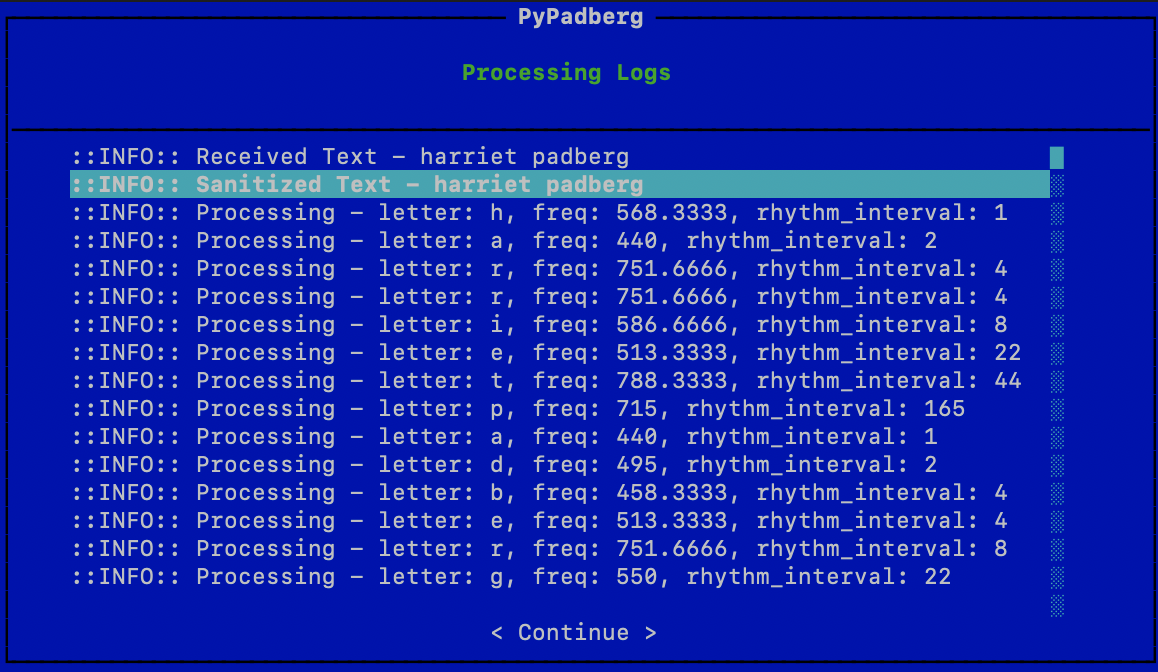}
	\caption{The processing screen in the PyPadberg Application, displaying how characters are mapped to frequencies and rhythms.}
	\label{fig:padScreenshot}
\end{figure}

\section{Padberg Repurposed}

Here we describe three compositions created using PyPadberg that demonstrate approaches to adapting a historical work for new contexts. \textit{Not Even One}, submitted by \anonymize{Molly Jones}, is an example of repurposing the historical and social context of the work by using the recreated composition program in a piece of music about the representation of women in the industry of music technology. \textit{Brevity} by \anonymize{Anna Savery} represents the reuse of an older work as a tool for composition with a performer. \textit{The Padberg Piano} by \anonymize{Anthony Caulkins} takes the concept of automatic composition from the original work in a new direction by creating an interactive software instrument.

\subsection{Not Even One}
\textit{Not Even One}, by \anonymize{Molly Jones}, was created as a submission to the 2nd Iannis Xenakis International Electronic Music Competition \footnote{http://xenakis.web.auth.gr/rules/, last accessed: 19th Apr 2019}. After discovering that the panel of judges selected for the competition consisted entirely of men, Jones decided to use the PyPadberg program to express the idea that women are a significant and under-appreciated part of the field of electronic music, and thus the presence of gender diversity in this judging panel is not only important, but necessary.  

After being made aware of Harriet Padberg via an invitation to create a composition using the PyPadberg program, \anonymize{Jones} discovered the surprising lack of information available about her despite Padberg's important place in the history of electronic and computer music - As mentioned, Padberg authored the first dissertation on algorithmic computer composition, and was a pioneer in the field of music therapy. A Google search of ``Harriet Padberg'' yields few related results, the top listed of which are obituaries, or local news articles written in response to her death. She does not have a Wikipedia entry. 

\anonymize{Jones} felt that Padberg's obscurity was a fitting representation of how, despite the long standing presence of women in, and the contributions of women to electro-acoustic music, the inclusion of women in organizations and events related to electro-acoustic music is seen as a novelty; if the perception is that women are new, or outsiders to the field, then their presence is non-required. Thus, the use of PyPadberg as a tool in the composition is a perfect thematic fit. 

\anonymize{Jones} began by feeding in three text patterns into the PyPadberg program and rending their corresponding melodies using a recording of herself speaking the phrase, ``The Xenakis Competition has no Women Judges,'' a polyphonic synthesizer, and the prepackaged violin sample. The text patterns themselves are not semantically meaningful, but were developed through trial and error for the melodies they created through the program. She then saved three and four voice canon/fugue versions of each pattern as wav files, and arranged them in Ableton Live \footnote{https://www.ableton.com/en/, last accessed: 19th Apr 2019}. The composition was built primarily around the four voice canon/fugue version of one of the three patterns rendered with the voice recording, with pitched and shifted versions of the other renderings used for ornamentation, accent and layering. 

The creation and submission of this piece was not meant to be seen as an act of aggression, or of mockery. \anonymize{Jones} hoped only to provide a reminder to the Competition of the importance of gender diversity and female inclusion, and desires that it instead be taken as a constructive criticism. 

On the creation of this piece and the use of the PyPadberg program, \anonymize{Jones} remarked, ``It is an ongoing duty to challenge the exclusion of innovative voices from present and past narratives, and I am happy to see that duty being carried out in the form of the PyPadberg program.''

\subsubsection{Author's Notes on Not Even One}
It is worth noting that PyPadberg was created without intention to frame the work as an attempt to change gender imbalances in representation. The work was created solely to recreate one of the first algorithmic computer compositions for the benefit of the authors and future users, and not as an attempt to benefit the system's creator. 

\subsection{Brevity}
\textit{Brevity}, by \anonymize{Anna Savery}, is a audio-visual piece which draws heavily on the Ben Folds song \textit{Still Fighting It} using PyPadberg to create a melody from its lyrics which acts as a base for tribute and reinterpretation. 

\textit{Still Fighting It} was dedicated to Folds' son and describes Folds' feelings about fatherhood, childhood, and watching his son grow. The accompanying music video contains a great deal of home video footage of Folds' son as a young child, around the time of the song's writing, and all non-candid footage shot specifically for the music video was made to have a similar aesthetic. As such, it serves as a beautiful encapsulation of a particular time and focus in Folds' life and the life of his child. \anonymize{Savery} desired to create a similar tribute to her own young daughters, seven months old and three years old at the time of this writing, encapsulating their early childhoods and expressing her own feelings about them and about parenthood while paying respects to the original work by Folds. 

\anonymize{Savery} selected a set of personally meaningful lyrics from \textit{Still Fighting It} as inputs to PyPadberg in order to create a melody from which to build out the rest of her composition. After saving the melody data as a CSV file, she recorded herself playing the melody live on several instruments and sampled those recordings along with recordings of her children's voices and recordings of them playing with their toys for arrangement and manipulation in Ableton Live. For the visual component, she created a matching set of video snippets comprised of custom visuals and processed clips from her own home videos which was loaded into Max/MSP\footnote{https://cycling74.com/, 19th Apr 2019}.

In order to closely relate the music and visuals such that their interaction would express a narrative, albeit one abstract and deeply personal, \anonymize{Savery} co-manipulated them, simultaneously interpreting the same CSV data from PyPadberg as two separate but aligned sets of instructions. Live enabled her to make micro-tonal pitch manipulations that were faithful to the system specified by Padberg's algorithm. 

\anonymize{Savery} felt that using the PyPadberg program enabled her to capture and express the chaotic tendencies of childhood through the filter of Padberg's micro-tonal program. In her own words, ``Children are notoriously noisy, and this tendency is amplified by their ever expanding collection of acoustic and battery operated toys. I wanted to capture that chaos by playing with some of these toys and altering [their sounds] with the influence of PyPadberg.''

\subsection{The Padberg Piano}

The \textit{Padberg Piano}, a new instrument and compositional system by \anonymize{Anthony Caulkins}, was designed to build upon the algorithm and letter-pitch organization outlined by Padberg. While a standard piano has twelve keys within an octave, the \textit{Padberg Piano} takes 24 keys (or two regular octaves) to repeat a pitch, as shown in (Fig 2.). The term octave is chosen, despite not representing a standard scale, due to its use when describing repeated notes in different registers. 

\begin{figure}[]
	\centering
		\includegraphics[width=1\columnwidth]{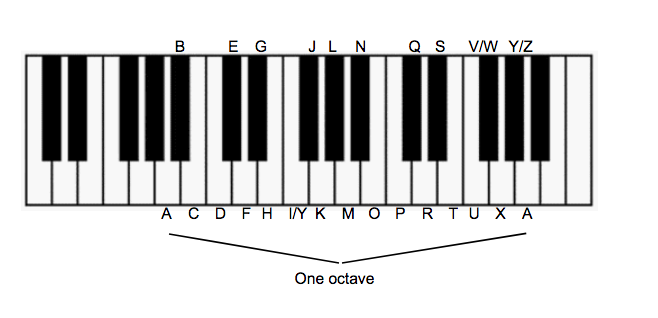}
	\caption{Padberg Keyboard}
	\label{fig:padkey}
\end{figure}

The \textit{Padberg Piano} is software instrument built within the Kontakt platform and can be accessed as a plugin within most audio software. As this instrument is designed with a piano keyboard in mind, it is possible to  compose music in standard notation to be played as if it were a traditional piano. However, many of the notes on the page will not correspond to the pitches coming from the piano. Thus the \textit{Padberg Piano} is a transposing instrument.

\subsubsection{Composing with The Padberg Piano}

To ease the difficulty of moving from a 12 pitch-class system to 24, \anonymize{Anthony Caulkins} has worked to create a systematic method of composition, drawing on algorithmic processes described by Padberg as well as Stephen Dembski's compositional system - Circles \cite{dembski2005}. Dembski's methods are drawn upon to design a pseudo-tonality  upon which to derive rules of 'harmony, 'consonance', and 'dissonance', as opposed to the contrapuntal nature of Padberg's original canons and 'free-fugues'.

\subsubsection{Proposal Of A ``New Tonal'' System}

The alphabetical ordering of Padberg's mappings as shown in Table 1 can be likened to the 12 pitch-class chromatic scale of the standard equal temperament system. In both, adjacent pitch-classes are of interval-class 1 (one semi-tone). In a 12 pitch-class set, interval-class 1 is inversionally symmetrical with interval-class 11, 2 with 10, 3 with 9, and so on (see Fig. 3), so an increase in pitch by interval-class 1 will result in the same pitch-class as a decrease by interval class 11. Similarly, in a 24 pitch-class set, interval-class 1 is inversionally symmetrical with 23, 2 with 22, etc. 
\begin{figure}[]
	\centering
		\includegraphics[width=1\columnwidth]{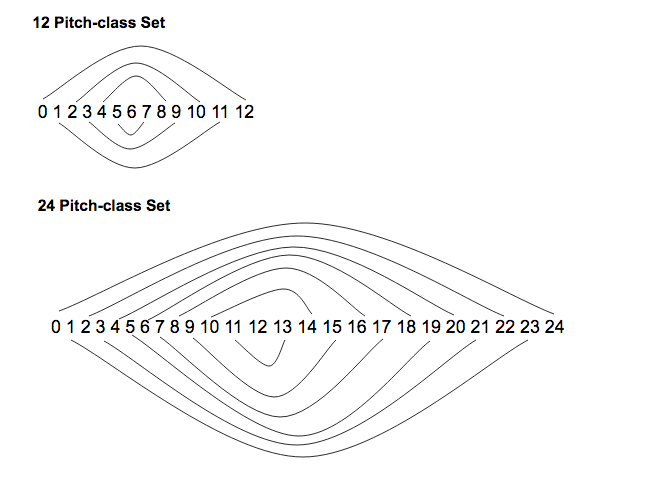}
	\caption{Inversionally Symmetrical Interval Classes}
	\label{fig:sym}
\end{figure}

Another possible ordering of a 12 pitch-class collection which accounts for each pitch-class is one with an adjacency interval-class of 5 (or 7 - its modulo 12 inversion). This ordering is also known as the Circle of Fourths/Fifths. Interval-classes 1 and 5 (or 11 and 9) are the only possible options to account for each of the 12 pitch-classes of the chromatic collection. An adjacency interval-class of 11 (or 13 - its modulo 24 inversion) is needed to scale the Circle of Fourths to the Padberg 24 pitch-class set. Interval-class 11 in the 24 set is analogous to interval-class 5 in the 12 set as they are both one step away from the interval-class at the point of symmetry (12 and 6 respectively). 

Next, two new orderings based on interval-class adjacency 5 (in the 12 pitch-class set) and 11 (in the 24 pitch-class set) are created. From here, seven adjacent numbers in the 12 pitch-class collection and 13 adjacent numbers in the 24 pitch-class collection are selected. The chosen letters are aligned over their positions on the original 'chromatic' collections. The circled letters in Fig. 4 represent the 'tonic' note of each collection.

\begin{figure}[]
	\centering
		\includegraphics[width=1\columnwidth]{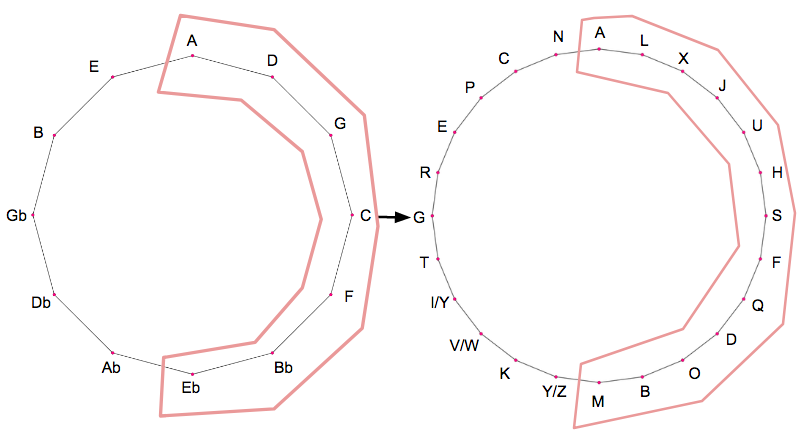}
	\caption{'Tonal' Collections}
	\label{fig:tonal2}
\end{figure}

At this point, the collections of pitches have been ordered into scales and related to a larger 'chromatic' collections. The circled letters in the 12 pitch-class (Fig. 4) collection makeup a Bb scale. Similarly, the Padberg alphabet is now organized into a 'tonal' system, complete with a transposable scale of whole and half steps, with the letter B as the 'tonic' (Fig. 4). 'Consonant' harmonies can be derived by connecting non-adjacent notes in each of the 'tonal' collections. In the 12 pitch-class set, the maximum number of connectable non-adjacent notes is 3 (which represents a tonal triad - Bb D F; C Eb G; D F A; etc.). The 24 pitch-class set will create basic chords of 5 notes, rather than 3 (see Fig. 5). 

\begin{figure}[]
	\centering
		\includegraphics[width=1\columnwidth]{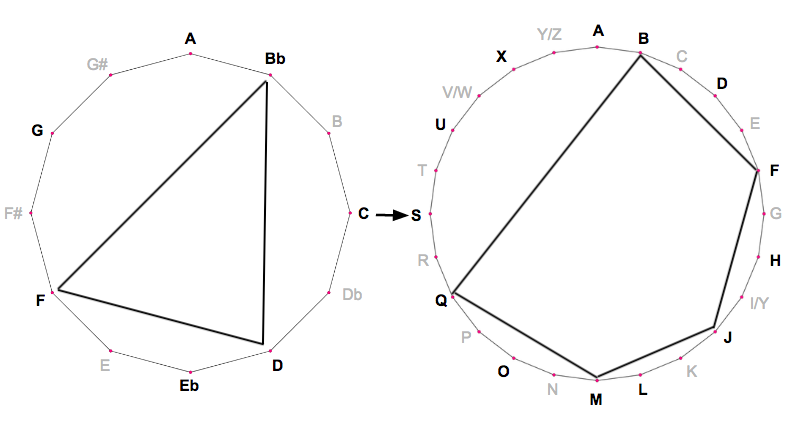}
	\caption{Non-Adjacent 'Consonances'}
	\label{fig:tonal6}
\end{figure}

5 notes are used rather than 6 (the maximum possible number of non-adjacent pitches) to maintain an the important tonal property of the gamut, whereby the tonic triad contains three notes, the dominant contains three notes (one overlapping the tonic), and the subdominant contains three notes (with one overlapping the tonic). These three functional harmonies represent all of the pitches of the scale. In the 24 pitch collection the tonic will be built on B (B F J M Q), the dominant on Q (Q U A D H), and the subdominant on L (L O S X B). 

Now that systems of scale, transposition, consonance, dissonance, and functional harmony have been developed, contrapuntal composition can begin. In the spirit of Padberg's canons, Caulkins has translated Bach's first keyboard invention to Padberg's 24 note system. Unsurprisingly, some difficulties arise when translating music designed for 12 pitch-classes to 24. Namely, how do scale degrees and functions move from a smaller system to a larger? Attempting to maintain the tonic, dominant, and subdominant functions outlined above, scale degree equivalency from Bach to Padberg is: 1 - 1, 2 - 2, 3 - 5, 4 - 6, 5 - 9, 6 - 11, 7 - 12. This might seem slightly confusing as there is a certain degree of arbitrariness in assigning scale degrees 2, 3, and 6. However, the aim is in maintaining the tonal functions of these scale degrees as well as possible. 

One critical difference is the range of the Bach/Padberg is much larger due to the fact that it takes 24 piano keys to cycle through a full octave. Also, in the Bach/Padberg score, each accidental is explicitly notated, rather carrying through a measure. The reason for this is that the 13 note scale cannot be represented by standard key signatures. 

\begin{figure}[]
	\centering
		\includegraphics[width=1\columnwidth]{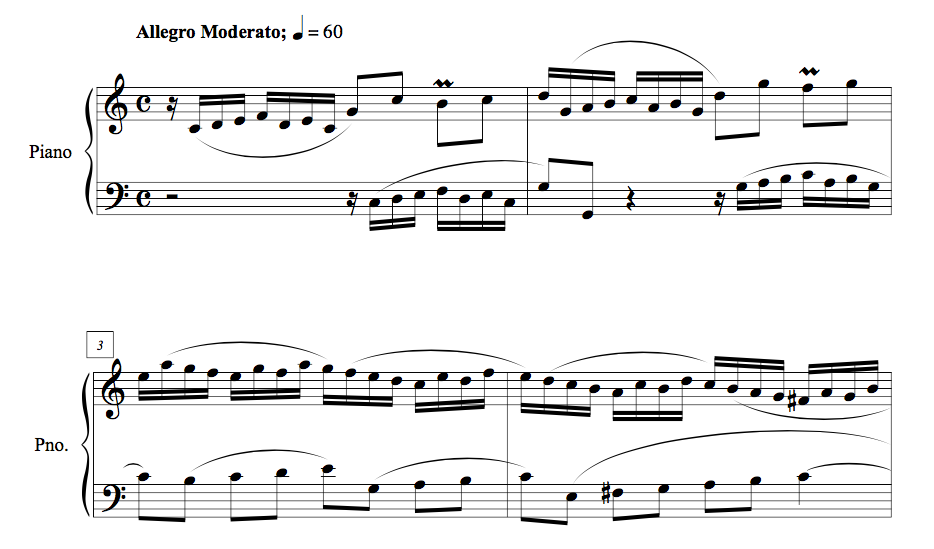}
	\caption{Original Bach}
	\label{fig:anthony}
\end{figure}

\begin{figure}[]
	\centering
		\includegraphics[width=1\columnwidth]{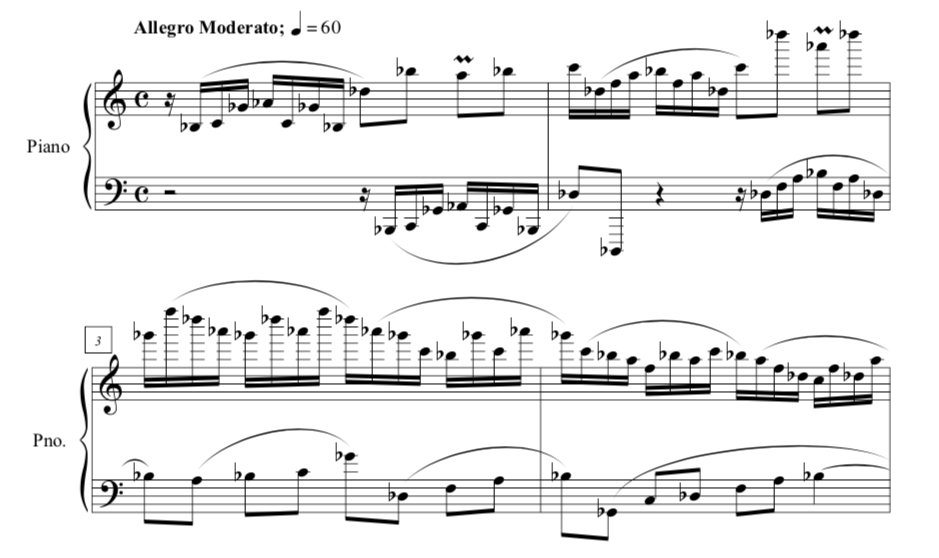}
	\caption{Padberg Piano}
	\label{fig:bach}
\end{figure}

\subsubsection{Reflections on Padberg Piano}
After completing the work, Caulkins reflected on using Padberg's System. ``Overall the success of the Padberg Piano is that it allows for free composition using Padberg's pitch system while giving a composer the freedom to employ ``text to music'' processes however they might wish. Obviously the sound-world of the piano is somewhat foreign to most music listeners, however when put into the context of a Bach invention, learning to hear functional shifts (chord changes, key changes, etc.) becomes fairly intuitive. Difficulties surrounding composing for this instrument become apparent within the context of re-purposing functional harmony into a 24 pitch-class system. Certain arbitrary decisions of course have to be made when scaling the standard tonal system into this new one. However, the product is something quite learn-able and fulfilling."

\section{Conclusion}

In this work, we discussed the obscure yet historically significant Harriet Padberg and her 1964 PhD dissertation \textit{Computer Composed Canon and Free Fugue}, and presented our own recreation or reinterpretation of her algorithm, PyPadberg. PyPadberg has been open sourced both to bring this algorithm into the present dialog on music technology history and for modern composers to use. Finally, and most critically, we've presented specific examples of how composers can use PyPadberg and other historical works as a basis or tool for new works. The three examples we present display the use of a PyPadberg for its surrounding narrative, as a means of producing a particular aesthetic, and as a basis for the re-formalization of aesthetic constructs such that one can filter other works through Padberg's aesthetic lens.

\section{Acknowledgments}
Many thanks to \anonymize{Jason Freeman} for providing the impetus for development and guidance during the development of PyPadberg.


%
%


\bibliographystyle{abbrv}
\bibliography{padberg}

\begin{thebibliography}{10}

\bibitem{rscjWebsite}
Harriet padberg, rscj | https://rscj.org/about/memoriam/harriet-padberg-rscj.

\bibitem{maryvilleUniversity}
Music therapy | https://www.maryville.edu/hp/music-therapy/.

\bibitem{ariza2011}
C.~Ariza.
\newblock Two pioneering projects from the early history of computer-aided
  algorithmic composition.
\newblock {\em Computer Music Journal}, 35(3):40--56, 2011.

\bibitem{bashe1986}
C.~J. Bashe, L.~R. Johnson, J.~H. Palmer, and E.~W. Pugh.
\newblock {\em IBM's Early Computers}.
\newblock MIT press, 1986.

\bibitem{dembski2005}
S.~Dembski.
\newblock An idea of order.
\newblock {\em Perspectives of New Music}, 43/44:403--424, 2005.

\bibitem{fernandez2013}
J.~D. Fern{\'a}ndez and F.~Vico.
\newblock Ai methods in algorithmic composition: A comprehensive survey.
\newblock {\em Journal of Artificial Intelligence Research}, 48:513--582, 2013.

\bibitem{hiller1968}
L.~Hiller.
\newblock {\em Music Composed with Computer[s]: A Historical Survey}.
\newblock Number~18. University of Illinois, 1968.

\bibitem{nilson2011}
C.~Nilson.
\newblock Dvd program notes.
\newblock {\em Computer Music Journal}, 35(4):119--137, 2011.

\bibitem{padberg1964}
H.~Padberg.
\newblock {\em Computer-Composed Canon and Free-Fugue}.
\newblock PhD thesis, St. Louis University, 1964.

\bibitem{tenYearsMaryville}
H.~A. Padberg.
\newblock Ten years of music therapy at maryville college and saint louis
  (december 10, 1982 speech).

\end{thebibliography}

\end{document}